\documentclass[conference,a4paper]{IEEEtran}
\IEEEoverridecommandlockouts
\usepackage[T1]{fontenc}
\usepackage{times}

\usepackage{amssymb,amsmath,amsfonts,amsthm}
\usepackage{stmaryrd}
\usepackage{xcolor}
\usepackage{footnote}
\usepackage{graphicx}
\usepackage{algpseudocode}
\usepackage[framemethod=latex]{mdframed}
\usepackage{caption}

\usepackage{enumerate}

\newcommand{\Fdn}{\ensuremath{\mathbb{F}_2^n}}

\newcommand{\Fdk}{\ensuremath{\mathbb{F}_2^{n-r}}}
\newcommand{\FdG}{\ensuremath{\mathbb{F}_2^{(n-r)\times n}}}
\newcommand{\FdH}{\ensuremath{\mathbb{F}_2^{r\times n}}}
\DeclareMathOperator{\wt}{wt}

\author{
\IEEEauthorblockN{Julia Chaulet}
\IEEEauthorblockA{Inria, Paris, France}
\IEEEauthorblockA{Thales TCS, Gennevilliers, France\\
\texttt{julia.chaulet@inria.fr}}
\and
\IEEEauthorblockN{Nicolas Sendrier}
\IEEEauthorblockA{Inria, Paris, France\\
\texttt{nicolas.sendrier@inria.fr}
}
\thanks{This work was supported in part by the Commission of the European 
Communities through the Horizon 2020 program under project number 645622 
PQCRYPTO.}
}

\date{}
\title{Worst case QC-MDPC decoder for McEliece cryptosystem}
\begin{document}
\maketitle
\begin{abstract}
  QC-MDPC-McEliece is a recent variant of the McEliece encryption
  scheme which enjoys relatively small key sizes as well as a security
  reduction to hard problems of coding theory. Furthermore, it
  remains secure against a quantum adversary and is very well suited
  to low cost implementations on embedded devices.

  Decoding MDPC codes is achieved with the (iterative) bit flipping
  algorithm, as for LDPC codes. Variable time decoders might leak some
  information on the code structure (that is on the sparse parity
  check equations) and must be avoided. A constant time decoder is
  easy to emulate, but its running time depends on the worst case
  rather than on the average case. So far implementations were focused
  on minimizing the average cost. We show that the tuning of the algorithm is not the same to reduce the maximal
  number of iterations  as for reducing the average cost. This provides some indications on how
  to engineer the QC-MDPC-McEliece scheme to resist a timing
  side-channel attack.
\end{abstract}

\section{Introduction}

With the advance of quantum computing, efficient algorithms against
number theory based cryptosystems \cite{S94a} have become a threat that
cannot be neglected. There is a need to develop post-quantum
cryptosystems, that is cryptosystems which remain secure against an
adversary equipped with a quantum computer.

Among all possible techniques for post-quantum cryptosystems we
found code-based cryptography and in particular the McEliece public-key
encryption scheme \cite{M78}. The original McEliece scheme uses
(randomly permuted) binary Goppa codes and so far has resisted all
cryptanalytic attempts even from quantum adversaries. The security is
twofold and relies on two assumptions, the hardness of generic
decoding, the {\em message security}, and the pseudorandomness of
Goppa codes, the {\em key security}. Generic decoding is NP-complete
\cite{BMT78} and is also believed to be hard on average. Though the
pseudorandomness of Goppa codes has not been studied as thoroughly as
generic decoding, no efficient algorithm is known to distinguish a
random matrix from a generator matrix of a Goppa code, except when the
code rate is close to one \cite{FGOPT11}, a case which is not a threat
against the McEliece encryption scheme.

One of the drawback of code-based public-key schemes is that they
require large public keys (a generator matrix of the code). Gaborit proposed to use
 quasi-cyclic codes \cite{G05} to solve
that problem. As demonstrated in \cite{S10}, the use of quasi-cyclic codes does not significantly change the security
reduction. However, the key security, which essentially
requires that the public key (a generator matrix) does not leak
information on the algebraic structure, is more problematic, and
quasi-cyclic codes with algebraic structure may sometimes have
vulnerabilities \cite{FOPT10}.

In 2013, the use of quasi-cyclic MDPC (Moderate Density
Parity Check) codes was suggested to instantiate the McEliece
scheme in  \cite{MTSB13}. This version of McEliece enjoys relatively small key sizes (a
few thousand bits) and its security provably reduces to the hardness
of deciding whether a given quasi-cyclic code contains a word of
small weight or not. Moreover, the decryption essentially consists in
decoding and can be achieved with the same iterative algorithms as for
LDPC codes. In particular, a low cost implementation, suitable for embedded systems, can be achieved using a hard decision bit flipping  iterative algorithm, as demonstrated in \cite{HMG13}.

\noindent{\em Our contributions.}
In this paper we will further examine the decoding of MDPC codes. The
bit flipping decoding algorithm involves a threshold which can be
chosen in many different ways and which may change along the
iterations of a given decoding instance.  A series of paper
\cite{HMG13,MG14a,MOG15} are focused on how to choose the threshold
in order to reduce the average number of iterations to successfully
decode a given number of errors. However, if an adversary is able to
measure, through a {\em side-channel}, the number of iterations of the
decoder, he might be able to deduce some information on
the secret key (the sparse parity check matrix).  That is to say, for instance,
 by submitting properly chosen noisy codewords to the decoder and
observing the decoder behavior for each of them.

Such an attack can be countered by designing a constant time
decoder. In the present case this means adding fake iterations
until we reach a prescribed number of iterations. This
number will be chosen such that stopping the decoding process at
 this point ensures a negligible probability of failure. In other
words, we wish to optimize the decoder
behavior in the worst case rather than in the average 
case in order to minimize the decoding cost.

The main observation of this paper is that minimizing the average cost
leads to significantly different algorithms than minimizing the worst
case. Using a heuristic approach we have determined a variant of the bit flipping algorithm which favors the worst case for a particular set of parameters. Combined with intensive simulation, it allows us to
provide some guidelines for the engineering of QC-MDPC-McEliece
decryption aiming to resist to above mentioned timing
attacks. It also allows us to understand how to choose the parameters
of both the system and the decoding algorithm such that the probability
of decoding failure remains negligible.

\section{Preliminaries}
$\mathbb{F}_2$ denotes the field with two elements, $\wt()$ denotes the
Hamming weight.
\subsection{Moderate Density Parity Check Codes}
The Moderate Density Parity Check (MDPC) code construction is very
similar to the LDPC code's one. They both are binary linear codes
defined by a sparse parity check matrix $H$ and only differ on the
sparseness of this matrix. For MDPC codes, the parity check matrix row
weight $w$ is much larger than for LDPC codes, typically
$w=O(\sqrt{n})$. We denote $(n, r, w)$-MDPC, a MDPC code of length
$n$, codimension $r$ whose sparse parity check matrix has a row weight
$w$.  \newtheorem{linear code}{Definition}
\begin{linear code}
  A square matrix is said \textit{circulant} if its rows are the
  successive cyclic shifts of its first one.
\end{linear code}
A linear code is \textit{quasi-cyclic} (QC) if its generator or parity
check matrix is composed of circulant blocks.

Though the parameters are flexible, all instances of QC-MDPC-McEliece
proposed in \cite{MTSB13} use codes of rate $1/2$.  For the sake of
simplicity, we will do the same in the rest of this paper and consider
QC-MDPC codes with two (circulant) blocks, $H =[H_0 \mid H_1] \in \FdH$.

\subsection{McEliece Cryptosystem Using (QC-)MDPC Codes}
Let $t$ denotes the number of errors which can be corrected by a bit
flipping iterative decoder of an $(n,r,w)$-MDPC code with
$n=2r$. Typically, we expect that $tw=O(n)$ and thus as
$w=O(\sqrt{n})$, we obtain $t=O(\sqrt{n})$. The McEliece
cryptosystem instantiated with (QC-)MDPC codes works as follow.
\begin{enumerate}
\item \textit{Key Generation.}  Generate a (two) block circulant
  parity check matrix $H=[H_0 \mid H_1]\in \FdH$ with rows of
  weight $w$ and such that $H_1$ is invertible. Compute its associated
  generator matrix $G=[I \mid (H_1^{-1}H_0)^T] \in \FdG$ in systematic form.  The
  \textit{public key} is $G$ and the \textit{private key }is $H$.
\item \textit{Encryption.}  Let $m \in \Fdk$ be the
  plaintext. Generate a random vector $e \in \Fdn$ such that $\wt(e) =
  t$. The \textit{ciphertext} is $c = mG + e \in \Fdn$.
\item \textit{Decryption.}  Decode the ciphertext $c$ to get a
  codeword $mG$. The \textit{plaintext} is obtained by truncating the
  first $n-r$ bits of that codeword.

\end{enumerate}

\subsection{Security Assessment}
Note that using LDPC codes for the McEliece scheme is deemed to be
unsafe \cite{MRS00}. However increasing the parity check matrix
density appears to have a positive effect in that respect.
\subsubsection{Reduction}
One of the strong feature of QC-MDPC-McEliece is its security
reduction. Following \cite{S10}, the system remains secure as long as:
\begin{enumerate}[(i)]
\item Decoding $t$ errors in a QC $[n,n-r]$ binary linear code is hard.
\item Deciding whether the code spanned by some block circulant
  $r\times n$ matrix (orthogonal to the public key) has minimum
  distance $\le w$ is hard.
\end{enumerate}
Both of these problems are NP-hard in the non cyclic case
\cite{BMT78,V97}. Their exact status is unknown in the circulant case,
however there is a consensus to say that cyclicity alone will not make
the problem easy. Note that the situation is very similar to
lattice-based cryptography: nobody believes that the cyclic or
``ring'' versions of the generic lattice problems are easy, even
though there is no completeness results.
\subsubsection{Practical Security}
In practice, the best known attacks are all based on information set
decoding and its variants \cite{P62,S88,MMT11,BJMM12,MO15}, either for
decoding $t$ errors in the code spanned by the public key or for
finding words of weight $w$ in its dual. Details can be found in
\cite{MTSB13}, but typical sizes are:
\begin{itemize}
\item $(n,r,w,t)=(9602,4801,90,84)$ for $80$ bits of security
\item $(n,r,w,t)=(19714,9857,142,134)$ for $128$ bits of security
\end{itemize}
using a $(n,r,w)$-QC-MDPC code correcting $t$ errors
($S$ bits of security means that all known attacks cost $\ge2^S$
elementary operations).

\section{Decoding QC-MDPC Codes}
\subsection{Decoding Algorithm}
\label{sec:decode}
It is proposed in \cite{MTSB12} to use a hard decision decoder derived
from the bit flipping algorithm introduced by Gallager \cite{G63}
(originally used as a LDPC decoder) to decode QC-MDPC when used to
instantiate the McEliece cryptosystem. In fact, any LDPC decoding
algorithms from \cite{G63} can be used as MDPC decoder as long as the
error correction capability matches with the chosen parameters. Note
that the solution proposed in \cite{MTSB12} has a slightly lower
capability compared to the original algorithm from \cite{G63}. However, the error
rates used in practice to instantiate the McEliece cryptosystem are
lower than those required in information theory. For
instance, to obtain $80$-bits of security, the parameters from \cite{MTSB13}
 ensures an error rate around $0.87~ 10^2$. This algorithm seems well 
 dedicated when implementing the cryptosystem in a constraint environment
  because it uses simple and fast operations. Actually, it already exists
   several embedded implementations of the cryptosystem. In \cite{MG14a},
  von Maurich and G{\"{u}}neysu achieved a very lightweight
implementation of the McEliece scheme using QC-MDPC codes and a very
high speed implementation of several decoding algorithms for QC-MDPC
codes are proposed in \cite{HMG13}.
% the authors of \cite{MOG15} already implemented this cryptosystem on FPGAs and AVR microcontroller (they also proposed and compared several hardware implementation of other decoding algorithms for MDPC codes). 

\begin{figure}[h!]
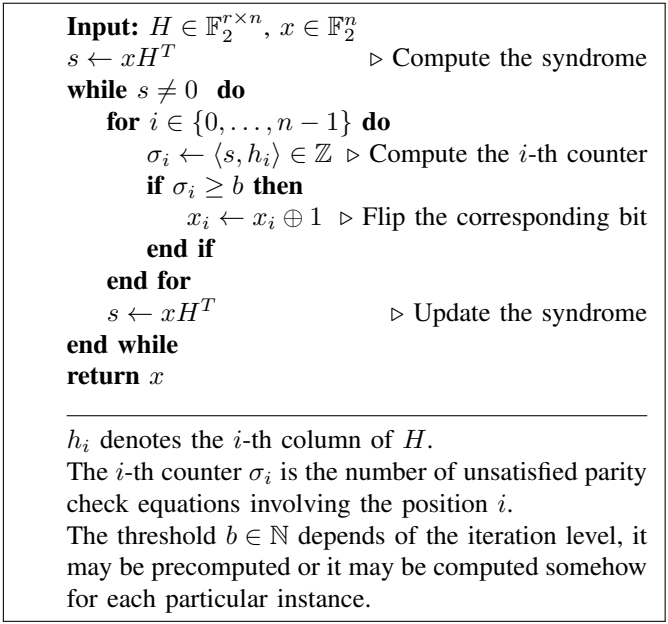

  % \begin{mdframed}
  \framebox{
    \begin{minipage}{0.93\linewidth}
      \begin{algorithmic}[C]
        \State \textbf{Input:} $H\in \FdH$, $x \in
        \mathbb{F}_2^n$

        \State $s \leftarrow  xH^T$ \Comment{Compute the syndrome}
        \While{$s \neq 0$ }
        \For{$i \in \{0, \ldots, n-1\}$}
        \State $\sigma_i \leftarrow \langle s, h_i \rangle \in \mathbb{Z}$
        \Comment{Compute the $i$-th counter}
        \If{$\sigma_i \geq b$}
        \State $x_i \leftarrow x_i \oplus 1$ \Comment{Flip the corresponding bit}
        \EndIf
        \EndFor
        \State $s \leftarrow xH^T$ \Comment{Update the syndrome}
        \EndWhile \\
        \Return{$x$}

        \State \hrulefill 

        \State $h_i$ denotes the $i$-th column of $H$.

        \State The $i$-th counter $\sigma_i$ is the number of unsatisfied
        parity check equations involving the position $i$.

        \State The threshold $b \in \mathbb{N}$ depends of the iteration level,
        it may be precomputed or it may be computed somehow for each particular instance.
      \end{algorithmic}
      \captionsetup{labelformat=empty}
      
    \end{minipage}

  }
  % \end{mdframed}
  \caption{Bit flipping algorithm}\label{fig:bitflip}
\end{figure}

The bit flipping algorithm is sketched in
Figure~\ref{fig:bitflip}. When $H$ is sparse enough and $x$ is close
enough to the code of parity check matrix $H$, the algorithm returns
the  closest codeword to $x$. The algorithm performance depends on the choice of $b$'s value.

In Gallager's original bit flipping algorithm, one threshold value per iteration
is precomputed. The formulas used to compute these
thresholds guarantee some probability of decoding failure. This
solution can be used to decode QC-MDPC codes but there is no certainty
regarding the probability of a decoding failure.

Another solution proposed in \cite{MTSB12} is to flip only bits that
violate the maximum number of parity check equations, up to a few units.

In \cite{HMG13} several other tunings were proposed in order to speed up
 the average running time of the algorithm. Instead of using the same
syndrome $s$ for all positions while an iteration, they proposed to
update the syndrome after each bit flip.

\subsection{Security Against Timing Attacks}
This algorithm is iterative and probabilistic, so we have to consider
two problems. First, the fact that the number of iterations of the
algorithm depends on both the error pattern and the parity check
matrix, may leak some information about the secret key and thus may
lead to a successful timing attack. For the moment, such attack is
unknown, but to avoid it we suggest to set the number of iterations of
the algorithm regardless of the instance. This number has to be large
enough to ensure a negligible decoding failure probability. These decoding failures
are the second problem to investigate. Somehow, a non zero probability of decoding
failure may also reveal information about the secret key. Obviously, 
considering a decoding failure as a decoding instance which require more iteration or
an infinite number of iteration, goes back to the first problem. That is 
to say, protecting the cryptosystem against timing attack.

 In practice, the average number of iterations to decode most of the
errors is very low, typically around 3 for a $(9602,4801,90)$-MDPC
code correcting 84 errors. However, a few error patterns require much
more iterations to be fully corrected. We will refer to these patterns
as \textit{worst cases} for the decoding algorithm. By abuse of
language, we will refer to the \textit{maximum number of iterations}
instead of the number of iterations observed in these worst cases.
 
Thus, our goal is to tune the algorithm to minimize the maximum
number of iteration rather than the average number of iterations
 and to find the better trade-off between overall running time and security.

\subsection{Tuning the Decoder}
We wish to find the best rules to tune the decoder, that is to say the best
rules for computing the threshold. Moreover, we want to keep a low
cost decoding procedure. Therefore the threshold computation may only
involve data which is easily obtained in the regular execution of the
bit flipping algorithm: the counter (number of unsatisfied parity
check equations) for each position, the syndrome (and also its
weight), the number of flipped bits  at each iteration, \ldots{}

In the next section, we describe our approach to optimize the
threshold choice.

\section{Minimizing the Number of Iterations for the Worst Cases}
\subsection{Approach}
As explained in the previous section, we want to minimize the maximum
number of iterations of the decoding algorithm in order to set this
value for each decoding instance. In section \ref{sec:decode},
we have highlighted that this algorithm relies on the counter values
for each position of the error vector, to decide whether the position
is considered as correct or wrong. Therefore, the threshold used to
make a decision is of main importance. Obviously, it has a direct
impact on the number of iterations needed: if the threshold is too
high, just a few errors will be corrected at each iterations, and if
the threshold is too low, the risk is to flip more correct bits (but
whose neighbors are false) than wrong bits.
%
% In the original Gallager's bit flipping algorithm, one threshold
% value is precomputed per iteration. The formulas used to compute
% these threshold guarantee a fixed probability of decoding
% failure. This solution can be used to decode QC-MDPC codes but there
% is no guarantee concerning the probability of a decoding
% failure. Nevertheless, it provides an efficient decoding algorithm.
%
% Another solution, proposed by Misoczki et al. \cite{MTSB12} is to
% flip bits violating the maximum number of parity check equations, or
% a few less. We observed that this technique ensures to decode any
% noisy codewords but using a high maximal number of iterations. In
% other words, this technique is sure but not very efficient.

In order to minimize the maximal number of iterations, we propose to
benefit from information available while decoding in order to avoid 
extra operations. As shown above, only few values are computed during
any of the iteration: the syndrome (and its weight), the  number of
unsatisfied parity check equations for each position and the number
of flipped bits. The syndrome weight appears to be a more relevant
information about how well the decoding is processed, because it only 
depends on the error vector and the parity check matrix. Furthermore, 
one can note that the sum of the number of unsatisfied parity check 
equations is proportional to the syndrome weight.  Thus, our idea was 
to take advantage of this information to adjust the threshold value for each
iteration and each instance, and decrease the maximum number of iteration required.

Our approach was to investigate a large amount of decoding traces (for
each iteration, we stored the value of: the error weight, the syndrome weight, the number of
flipped bit, \ldots) and to analyze instances requiring the highest
number of iterations. Then, we tried to decrease this number for these worst
cases by adapting the value of thresholds.

Our task was to obtain a global optimization, which was in fact much
more challenging than trying to minimize the error weight one
iteration at a time. It appears that the best solution was not
necessarily the obvious one: it can happen that decreasing the number
of corrected bits in the first iterations, also decrease the error
weight at the last iteration. Furthermore, it reveals that the
behavior of the worst cases cannot be well predicted using the weight
of the syndrome, the number of unsatisfied parity check equation for
each position or the number of flipped bits. Still, these information
helped us to analyze the decoding process.

The result of these investigations is that computing the thresholds as
a function of the syndrome weight, regardless to the current
iteration, seems to be the better solution to achieve the lowest
 maximum number of iterations.

\subsection{Results}

\begin{figure}[h!]
\includegraphics[scale=0.55]{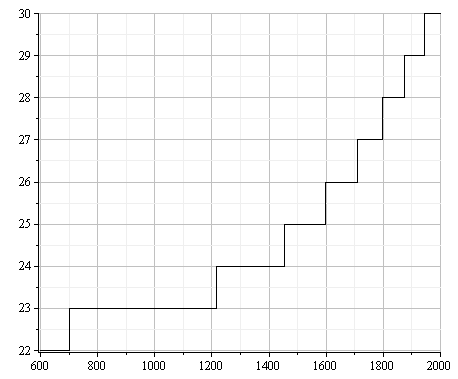}
\caption{Computation of the threshold values as a function of the
  syndrome weight for $(9602, 4801, 90)$-QC-MDPC codes in order to
  correct up to $84$ errors.}\label{fig:thrs}
\end{figure}

The Figure~\ref{fig:thrs} shows the step function used to compute
the threshold according to the syndrome weight that appears to be
the best choice in order to decrease the maximum number of iterations. 

As mentioned in \cite{MOG15}, the minimal average number of iteration
is achieved by updating the syndrome after each bit flip and by using
precomputed thresholds. However, it is also noticed that the higher
performance decoding algorithm can diverge for some particular instances. That is to say, some error
patterns cannot be decoded at all, and a second decoding attempt has 
to be done, using different threshold values. This does not affect much the
average running time of decoding but does not permit the design of 
a (reasonable) constant time decoder.

We compare in Table~\ref{tab:comp} the behavior of our worst case
decoder and the most efficient (on average) algorithm from
\cite{HMG13}.

\begin{savenotes}
\begin{table}[ht]
\centering
\begin{tabular}{|c|l|l|}
\hline
Iteration & This work & \cite{HMG13} algorithm\\
\hline
$2$ & $0.187 ~ 10^{-5}$ & $0.536$\\
$3$ & $0.768$ & $0.462$\\
$4$ & $0.231$ & $0.145~  10^{-2}$\\
$5$ & $0.408 ~  10^{-3}$ & $0.201 ~ 10^{-3}$\\
$6$ & $0.321 ~  10^{-5}$ & $0.152 ~ 10^{-5}$\\
$7$ & $0.150 ~  10^{-6}$ & $0.200 ~ 10^{-7}$\\
$\infty$\footnote{Proportion of errors which cannot be decoded.} & $ 0$ & $0.948 ~ 10^{-5}$\\
\hline
\end{tabular}
\caption{Proportion of errors patterns which are decoded in a fixed
  number of iterations. We are comparing the result of our work and
  the new algorithm from \cite{HMG13} which used precomputed thresholds
  and abort iteration when the syndrome becomes zero. We used $(9602,
  4801, 90)$-QC-MDPC codes and errors of weight $84$. The simulations
  were made over $1000$ codes with $10^{5}$ error patterns per code for our
  work and $500$ codes with $10^{5}$ error patterns per code for the
  algorithm from \cite{HMG13} }
\label{tab:comp}
\end{table}
\end{savenotes}

Note that, we do not find cases of divergence using our decoding
technique. These results suggest that the adjustment of the threshold
improves the worst case decoding but decreases the efficiency of the
algorithm for average cases. However, our goal was to minimize the
maximum number of iterations, which is $7$ for these parameters and
not to speed up the average running time of the decoder.  
Although we did not find instances which are not decoded after the $7^{th}$ iterations, 
we cannot guarantee the decoding failure probability to be zero. Hence, it is
probably safer to fix the number of iterations to $8$ or $9$ on
practice. An other viable solution could be to slightly increase the
length and the dimension of the code, keeping the same code rate and
the same values for $w$ and $t$. This would improve the overall
efficiency of the decoder.

% \section{Probability of Decoding Failure}

% As explained in the previous section, the decoding algorithm for MDPC
% codes is probabilistic, and, even with a fixed error weight, there is
% always a chance that the decoder fails (unless the error weight is
% ridiculously and uselessly small). In a cryptographic context, when
% these codes are used to instantiate the McEliece cryptosystem, a
% decoding failure can be exploited to get information about the
% secret. These techniques are unknown for the moment, but have to be
% considered.

% A theoretical analysis of the decoding algorithm for more than 3 or 4
% iterations seems rather difficult. A heuristic approach through
% simulations is certainly possible but may not allow a strong enough
% guaranty for cryptographic needs. 

\section{Conclusion}

Obtaining a constant time decoder for QC-MDPC is a safeguard against
timing attacks on the QC-MDPC-McEliece scheme. We show here that the
best constant-time decoder requires a specific tuning of the bit
flipping algorithm which relies on the syndrome weight. 
A limitation of our approach is that a
significant effort is needed to find the optimal threshold rule for
every set of code parameters. We think however our work gives a
serious hint on how to find the thresholds for others code parameters.

Finding theoretical bounds or estimates of the decoding failure
probability seems to be a challenging task, but for cryptographic
purpose, it could be of interest to reduce that probability to a very
small amount, say $2^{-128}$, with some kind of guarantee. We hope
that our preliminary work could open the way to such an achievement.

%\bibliographystyle{IEEEtran}
%\bibliography{codecrypto}

\end{document}